\documentclass[slac_one]{revtex4}
\usepackage{graphicx}
\usepackage{fancyhdr}
\def\met{\mbox{\ensuremath{\, \slash\kern-.6emE_{T}}}}

\begin{document}

\title{{\small{Hadron Collider Physics Symposium (HCP2008),
Galena, Illinois, USA}}\\ %% Please keep this conference title here
\vspace{12pt}
The ATLAS trigger menu for early data-taking} %% Paper title goes here


\author{T. Kono\\ for the ATLAS Collaboration}
\affiliation{CERN, Geneva 23, CH-1211, Switzerland}

\begin{abstract}
The ATLAS trigger system is based on three levels of event selection 
that select the physics of interest from an initial bunch-crossing rate 
of 40 MHz. During nominal LHC operations at a luminosity of 
$10^{34}$~cm$^{-2}$s$^{-1}$, 
decisions must be taken every 25 ns with each bunch crossing containing 
about 23 interactions. The selections in the three trigger levels must provide
sufficient rejection to reduce the rate down to 200 Hz, compatible with
the offline computing power and storage capacity. The LHC is expected to begin
operations in summer 2008 with a peak luminosity of 
$10^{31}$~cm$^{-2}$s$^{-1}$ with far fewer bunches than nominal running, 
but quickly ramp up to higher luminosities. 
Hence, we need to deploy trigger selections that can adapt to the changing 
beam conditions preserving the interesting physics and detector requirements 
that may vary with these conditions. 

We present the status of the preparation of the trigger menu for the 
early data-taking showing how we plan to deploy the trigger system from the
first collision to the nominal luminosity. We also show expected rates and 
physics performance obtained from simulated data.

\end{abstract}

\maketitle

\thispagestyle{fancy}


\section{Introduction}

The ATLAS trigger~\cite{hlt_tdr} is based on three levels of event selection:
Level-1 (L1), which is based on hardware and Level-2 (L2) and Event Filter (EF)
(collectively referred as the High Level Trigger or HLT) which are based on 
software algorithms analyzing the data on large computing farms.
The three levels of the ATLAS trigger system must reduce the output event
storage rate to $\sim$200~Hz from an initial LHC bunch-crossing rate of 
40~MHz. Large reduction against QCD processes is needed while maintaining 
high efficiency for low-cross-section physics processes including searches
for new physics~\cite{physics_tdr,detector_paper}.

Installation and commissioning of the LHC and ATLAS are currently going on 
and the LHC beam operation is expected to start in summer 2008. 
The commissioning of the ATLAS detector will begin with single-beam
operation and then with colliding beams at low luminosity with only 
a few bunches, but quickly ramp up towards the LHC design luminosity.
During the commissioning phase and in early physics runs, therefore, it is 
necessary to prepare appropriate triggers at each stage, in order to 
collect useful data for understanding of the ATLAS detector, the 
performance of the reconstruction software and the trigger selections.

At each trigger level, several selection logics can be implemented, where
the final decision is taken as the OR of all logics. The entire configuration
of the selection criteria used at the three levels is called the
trigger menu. 
The trigger menu must be defined taking into account the availability
of detector sub-systems at each phase of the commissioning, 
the efficiency of selecting signal events and the output rate at each level.
Currently, several trigger menus for the startup and early physics runs
have been developed and the performance was studied with Monte Carlo (MC) 
simulation.

\section{The ATLAS trigger system}
The ATLAS trigger consists of three levels (L1, L2 and EF) and 
selects events of physics interest. The event rate must be reduced 
from the initial bunch-crossing rate of 40~MHz 
to 75~kHz, 1~kHz and 200~Hz after L1, L2 and EF, respectively 

The level-1 trigger is a hardware trigger using data from the calorimeters and 
muon trigger chambers. The L1 calorimeter trigger is capable of triggering
on different types of objects: electromagnetic clusters (EM), hadronic clusters
(TAU), jets (JET), forward jets (FJ), missing transverse energy (XE),
total transverse energy sum (TE) and total jet transverse energy sum (JE).
The L1 central trigger processor (CTP) receives the multiplicities of 
localized objects above programmable thresholds (e.g. EM clusters with $E_T$
greater than 20~GeV) found by the calorimeter and muon sub-systems, 
as well as threshold information for the global quantities (XE, TE and JE). 
The CTP makes a decision 
based on multiplicities of various thresholds. 

While the threshold values
and the logic used in the CTP are configurable, the number of different 
thresholds for each type of object and the number of independent 
trigger logics (trigger item) are fixed by the hardware design.
In particular, the maximum number of trigger items is 256.

The level-2 trigger is a software-based trigger running on $\sim$500
computing nodes with quad-core dual CPU on each node. 
The L2 system receives the results of the L1 trigger
items and information on Regions of Interest (RoI) where L1 observed 
interesting objects. Execution of L2 algorithms is controlled by 
the HLT Steering software 
which runs the algorithm for each RoI. Algorithms then request data 
from the detector readout system belonging to the RoI and process
data in this RoI. This reduces the time spent on the data transfer and the
data processing, allowing the execution time per event to be below 40~ms.
The EF execution is also controlled by the HLT Steering. 
At EF, algorithms similar to the ones in the offline reconstruction are run.
Since EF algorithms run after the event building, full event data are 
available.

Events are written out into one or several Streams depending on the EF
chains that were passed. 
The data streaming allows one to reprocess events separately 
for each stream. This would be useful, for example, if new calibration
constants for the muon spectrometer become available,
one can reprocess events in the muon stream first without 
processing events triggered only by the jet trigger.
The ATLAS experiment has adopted the inclusive streaming model, 
i.e. an event passing several EF chains may be directed to several streams. 
This makes the system simpler at the analysis stage since all events
passing the same trigger are contained in the same stream. 
However, this results in duplicating event data, 
so the overlap between streams must be kept as low as possible.

It is planned to have five to ten physics streams. 
In addition to physics streams,
there are two special types of streams. One is the express stream which 
will be used for prompt reconstruction with a short delay after 
the data-taking.
The primary purpose of the express stream is monitoring and debuging before 
the bulk reconstruction, therefore events in the express stream consist of 
a sample of high-purity signal events from the physics streams.
The other type of stream is the calibration stream which contains events
triggered by calibration triggers used to collect a large data sample 
for detector calibrations. 
Events which caused errors during online running, e.g. time-outs, crashes etc.
are sent to the debug stream to be investigated further.

\section{Trigger menu for early running}

\subsection{Strategy of the trigger menu commissioning}
At the beginning of LHC beam commissioning, the luminosity is expected to 
be very low ($L<10^{31}$~cm$^{-2}$s$^{-1}$) with only a few bunches in the
LHC ring. 
During the low-luminosity running, rates of physics process are rather
low and emphasis will be put on commissioning the trigger system and 
understanding the trigger performance for higher luminosities.

We plan to start with a simple menu to first trigger on filled bunches
using the beam pick-up system (BPTX) and Minimum Bias Trigger Scintillators
(MBTS). Data collected by such triggers can be used to adjust the timing
of various detector system with respect to the LHC bunch-crossing. 
Once the timing-in of the sub-systems has been done, the standard L1 trigger
can be turned on at low thresholds, but initially running the HLT 
in pass-through mode which means that HLT algorithms are run, but not used 
to make the trigger decision. 
These data can be used to commission the HLT selection.

\subsection{Trigger menu for L=$10^{31}$~cm$^{-2}$s$^{-1}$}

\begin{table}[htbp]
  \begin{center}
    \begin{tabular}{l|c|c}
      \hline
      Signature & Lowest L1 threshold (GeV) & Lowest unprescaled L1 threshold (GeV) \\
      \hline
      Muon & 4 & 4 \\
      EM cluster & 3 & 7 \\
      TAU (hadronic cluster) & 6 & 40 \\
      Jet & 10 & 120 \\
      $\met$ & 15 & 70 \\
      $\Sigma E_T$ & 150 & 650 \\
      $\Sigma E_T^{jet}$ & 120 & 340 \\
      \hline
    \end{tabular}
    \caption{Lowest L1 $E_T$ thresholds used in the menu at the luminosity of 
      $10^{31}$~cm$^{-2}$s$^{-1}$.}
    \label{tab:thresholds}
  \end{center}
\end{table}

\begin{figure}[htbp]
  \begin{center}
    \includegraphics[width=0.6\linewidth]{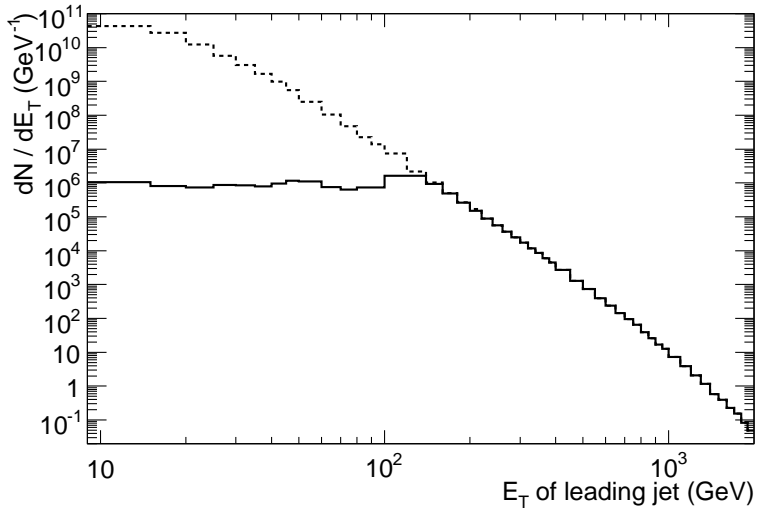}
    \caption{Jet rate spectrum as a function of $E_{T}$.}
    \label{fig:jet_rates}
  \end{center}
\end{figure}

For early physics runs, several trigger menus for various luminosities
are being prepared. In the startup phase, the  luminosity is expected to be 
much lower than the LHC design luminosity ($10^{34}$~cm$^{-2}$s$^{-1}$).
We have prepared a comprehensive menu to be used at a luminosity of 
$10^{31}$~cm$^{-2}$s$^{-1}$, which includes about 130 L1 items 
and $\sim 180$ L2 and EF chains.
At this luminosity, the physics emphasis is to trigger on Standard Model
processes such as QCD jets, $B$-physics and $W$/$Z$ events and 
for use in understanding the detector and reconstruction
performance.
For the trigger itself, data collected at low luminosity should allow studies
of the trigger selection algorithms and the optimization of the trigger
for higher luminosities.

\begin{figure}[htbp]
  \begin{center}
    \includegraphics[width=0.6\linewidth]{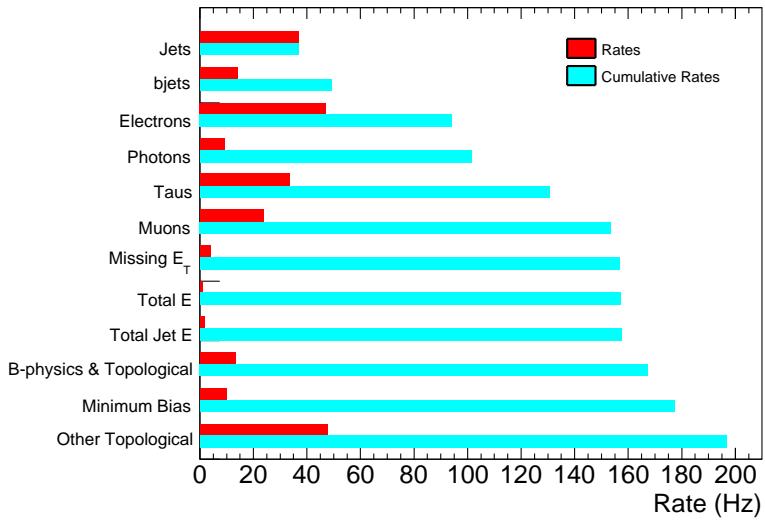}
    \caption{EF output rates for each stream.}
    \label{fig:L31_rates}
  \end{center}
\end{figure}

At $10^{31}$~cm$^{-2}$s$^{-1}$, rates are comparatively modest,
so low thresholds and loose selections can be used at L1, and the HLT can be
run in pass-through mode for most physics processes.

Table~\ref{tab:thresholds} shows the lowest thresholds used for each L1 
object type with and without prescales. 
At L1, all muon thresholds can be run unprescaled. 
They are used for single and multi-muon triggers, 
and also for $B$-physics triggers at the HLT.
The lowest thresholds for unprescaled L1 items for EM and TAU cluster triggers
are 7 and 40~GeV, respectively, which run without any isolation criteria.
The EM object triggers are used for electron and photon triggers at the HLT.
The TAU trigger is used to trigger on the hadronic $\tau$ decay from
$Z$, $W$ and top. 
Eight L1 jet thresholds and corresponding prescales are chosen to give
an approximately flat trigger rate across the jet $E_T$ spectrum up to 
$E_T>100$~GeV, beyond which no prescaling is applied. 
The jet triggers are run in pass-through mode at the HLT at this luminosity.
The samples selected by jet triggers are used for QCD background studies 
for many channels. Figure~\ref{fig:jet_rates} shows the flat
jet $E_T$ spectrum as selected by the trigger. The total transverse energy and
$\met$ triggers are not expected to be crucial for the early running,
but are included in the menu to test the performance.

Although many triggers can be run with low thresholds without prescales,
triggers with higher thresholds, which can be used at the design luminosity,
are also included in the menu for test purpose.

\begin{table}[htbp]
  \begin{center}
    \begin{tabular}{l|c|c}
      \hline
      Stream & Total rate (Hz) & Unique rate (Hz) \\
      \hline
      egamma & 55 & 48 \\
      muon & 35 & 29 \\
      jetTauEtmiss & 104 & 89 \\
      minbias & 10 & 10 \\
      \hline
      express & 18 & 0 \\
      calibration & 15 & 13\\
      \hline
    \end{tabular}
  \end{center}
  \caption{Total and unique rates for a selected raw data stream 
    configuration.}
  \label{tab:streams}
\end{table}

In addition to single object triggers, there are a number of combined 
triggers requiring more than one type of signature, for example, 
$\tau+\met$, $\tau+e$, etc.
Combined triggers have low rates at the luminosity of 
$10^{31}$~cm$^{-2}$s$^{-1}$, but will become essential at higher luminosity.

Figure~\ref{fig:L31_rates} shows a summary of the EF output rates 
for different groups of triggers, estimated by running the trigger on 
7 million simulated minimum-bias events. Grouping of triggers 
is done in such a way that the ``Electron'' group consists of 
all single-electron and multi-electron
triggers. Combined signatures like $e+\mu$ are assigned to the
``Other topological'' group.
The rate of each group accounts for the overlap between individual
trigger chains belonging to the group. The cummulative rates account for 
the overlap between the trigger groups.
The total output rate of the EF is about 200~Hz, compatible with the
target output rate. The L1 and L2 rates are estimated to be about
12~kHz and 620~Hz, respectively.

A possible streaming configuration at a luminosity of 
$10^{31}$~cm$^{-2}$s$^{-1}$ is four physics streams: electrons and photons, 
muons, minimum-bias triggers and jet/$\tau$/$\met$ triggers.
The stream names indicate the type of trigger signatures
they contain. The streaming configuration is defined to have approximately the
same proportion of events in each stream and 
keep the total overlap to less than 10~\%.
The total and unique rates for this streaming configuration are shown 
in Table~\ref{tab:streams}. The overlap between muon triggers and $B$-physics
triggers is about 15~\%, so these have to be merged into one stream.
The same applies to the overlap between jet triggers and $\tau$/$\met$
triggers. The final optimization of the streaming configuration must be
done by analyzing the overlaps with real data.

\subsection{Evolution to higher luminosities}
As the luminosity increases, it becomes necessary to use higher thresholds 
and tighter selections, and also to turn on HLT selections.
Already at a luminosity of $10^{33}$~cm$^{-2}$s$^{-1}$, rates from
interesting physics events such as $W$ and $Z$ production become significant,
and the use of isolation criteria and $\met$ triggers become important.
Many prescaled triggers with loose selections and pass-through triggers 
are included in the menu to study the performance of tighter selections
and to optimize the menu for higher luminosities. The optimization must
be done by investigating the trigger performance on real data.

\section{Conclusion}

Several trigger menus for different phase of the commissioning and 
the early running of the ALTAS experiment are being developed. Initially,
the commissioning of the trigger system will be done using the BPTX
and MBTS as inputs to the L1 trigger. This allows one to adjust the timing
of various sub-systems of the ATLAS detector and to commission other L1
triggers. 
At low luminosity, e.g. $10^{31}$~cm$^{-2}$s$^{-1}$, it is possible to 
use low thresholds and to use only the L1 trigger to select events. 
This allows one to study the
performance of higher-threshold triggers and to optimize the HLT selections 
with real data, which is crucial for optimizing the trigger menu as
the luminosity increases.
The trigger menu for the luminosity of $10^{31}$~cm$^{-2}$s$^{-1}$ has been
developed and studied in detail using simulated minimum-bias events. 
Thresholds, prescales and streaming configurations have been optimized 
after a few iterations to meet the restriction of 200~Hz output.
For the menu at higher luminosites, triggers with higher thresholds
and combined signatures will be used for the main physics selections.
The initial running will allow further optimization of the trigger
menu and reliable extrapolation to higher luminosities.

\end{document}